\documentclass[a4paper,11pt]{article}
\usepackage{pos}

\title{Multiplicative NLO matching in DIS with Pythia 8}
\author*[a,b]{Ilkka Helenius}
\author[a,b]{Joni O. Laulainen}
\author[c]{Christian T. Preuss}

\affiliation[a]{University of Jyvaskyla, Department of Physics, P.O. Box 35, 40014 University of Jyvaskyla, Finland}

\affiliation[b]{Helsinki Institute of Physics, P.O. Box 64, 00014 University of Helsinki, Finland}

\affiliation[c]{Institut f{\"u}r Theoretische Teilchenphysik und Kosmologie, RWTH Aachen University, 52056 Aachen, Germany}

\emailAdd{ilkka.m.helenius@jyu.fi}
\emailAdd{joni.o.laulainen@jyu.fi}
\emailAdd{preuss@physik.rwth-aachen.de}

\abstract{We present a new multiplicative next-to-leading order (NLO) matching scheme in \textsc{Pythia~8} for deep-inelastic scattering (DIS) processes in electron-proton collisions. The implementation is based on numerical event-by-event integration over the virtual corrections and real-emission phase-space and matrix-element corrections (MECs) for the first parton-shower emission. The latter has been included for two parton-shower options currently available in \textsc{Pythia~8} to handle DIS processes, the default shower with \textsc{dipole-recoil} option and \textsc{Vincia}. The implementation has been validated against other available NLO-capable event generators and compared to experimental data from the HERA collider. In all cases we find very good agreement confirming the validity of the new \textsc{Pythia} implementation. We also study the impact of the MECs in each shower and consider the implications for the upcoming Electron-Ion collider.}

\FullConference{
}

\begin{document}
\maketitle

\section{Introduction}

\textsc{Pythia~8} \cite{Bierlich:2022pfr} is a general purpose Monte Carlo event generator capable of simulating high-energy particle collisions with several beam combinations. In electron-proton collisions the events can be classified in terms of virtuality of the exchanged boson, $Q^2$. Low-virtuality events can be described in the photoproduction framework where the flux of incoming photons can be factorized from the hard scattering, which then takes place between a photon or the partonic constituents of it. The high-virtuality events, in turn, can be described as deep-inelastic scattering (DIS) where a lepton scatters from a single parton in the proton target. The partons participating in the hard scattering are then evolved with the parton showers to trace back all the QCD emissions that take place after or before the hard scattering. This multi-parton final-state is then converted into a hadronic state by applying the string hadronization model including remnants of the beam particle.

Here, we present a new implementation of multiplicative next-to-leading order (NLO) matching in \textsc{Pythia~8} for neutral-current DIS \cite{Helenius:2026odp}. In order to achieve NLO accuracy we need NLO $K$-factors at the Born-level kinematics and matrix-element corrections for the first parton-shower emission. Similar techniques have been earlier considered for various proton-proton collision processes and recently a few implementations have also considered DIS in \textsc{Herwig} \cite{Platzer:2011bc,DErrico:2011wfa}, \textsc{Sherpa} \cite{Hoche:2018gti, Meinzinger:2025pam}, \textsc{Powheg-Box} \cite{Banfi:2023mhz, Borsa:2024rmh} and within the \textsc{PanScales} project \cite{vanBeekveld:2023chs}. Thus the novelty of the work is being the first internal NLO-matched setup within \textsc{Pythia~8} that can handle processes taking place in hadronic collisions. Another advance of the presented setup is that by performing each event generation step in a single framework ensures that internal variables, such as shower evolution scale, remain consistent throughout the whole simulation. Improved simulation capabilities for electron-hadron collisions will be useful for the upcoming Electron-Ion Collider (EIC) that is currently being built in Brookhaven National Laboratory in the US \cite{AbdulKhalek:2021gbh}.

\section{Framework}

\begin{figure}[htb]
\begin{center}
\includegraphics[width=0.5\textwidth]{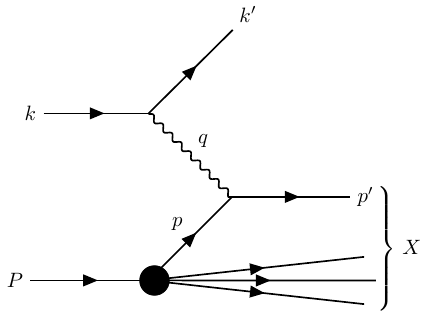}
\caption{Schematic illustration of DIS process in electron-proton collision and definitions of relevant momenta.}
\label{fig:DISkin}
\end{center}
\end{figure}

The cross section for DIS processes, shown in figure \ref{fig:DISkin}, can be written in terms of structure functions $F_1(x,Q^2)$, $F_2(x,Q^2)$ and $F_3(x,Q^2)$ \cite{ParticleDataGroup:2024cfk}. In the collinear factorization framework \cite{Collins:1989gx} one can write these as a convolution between the parton distribution functions (PDFs) and perturbatively calculable coefficient functions \cite{Altarelli:1979ub, Furmanski:1981cw}. To calculate the necessary NLO $K$-factors we have implemented a numerical integration to calculate the convolution for each event with sampled values for DIS invariants $Q^2 = -q^2 = -(k-k')^2$ and $x=\frac{Q^2}{2 P \cdot q}$ originally generated according to the leading order (LO) expressions. This convolution essentially accounts for integrating over the real-emission phase-space and, at the same time, combines this with the virtual corrections necessary to cancel out the divergences within these two contributions. The NLO weight is calculated and accounted for in the accumulated cross section calculation when the flag \texttt{SigmaProcess:reweightNLO} is switched on. To maximize the efficiency of this calculation we have so far considered only flavor-summed observables and the flavor selection is based on LO cross sections.

In addition to the NLO weight above, the first emission generated by the parton shower has to be corrected with the appropriate matrix element to make sure that real-emission rates match with the NLO calculation. We take the matrix elements for $\gamma^* q \rightarrow gq$ and $\gamma^* g \rightarrow q\bar{q}$ scatterings from Ref.~\cite{Catani:1996vz}. In DIS it is important to make sure that the parton shower algorithm does not distribute any recoil from the QCD emissions to the scattered lepton. In case of \textsc{Pythia~8} there are currently two options available, the \textsc{dipole-recoil} option within the default Simple Shower \cite{Sjostrand:2004ef, Cabouat:2017rzi}, and the \textsc{Vincia} antenna shower \cite{Brooks:2020upa}. The resulting matrix element corrections (MECs) for each of the considered parton showers are derived in Ref.~\cite{Helenius:2026odp}. We notice that the MECs can reach up to values of 2 at certain kinematical points. To account for the enhanced emission rates, the parton-shower acceptance ratios are increased by this value when MECs are applied. From \textsc{Pythia} version 8.317 on, these corrections are accounted for by default for the first emission. The release also contains an example \texttt{main341.cc} providing full list of recommended settings for NLO-matched predictions.

\section{Validation}

\begin{figure}[htb]
\begin{center}
\includegraphics[width=0.48\textwidth]{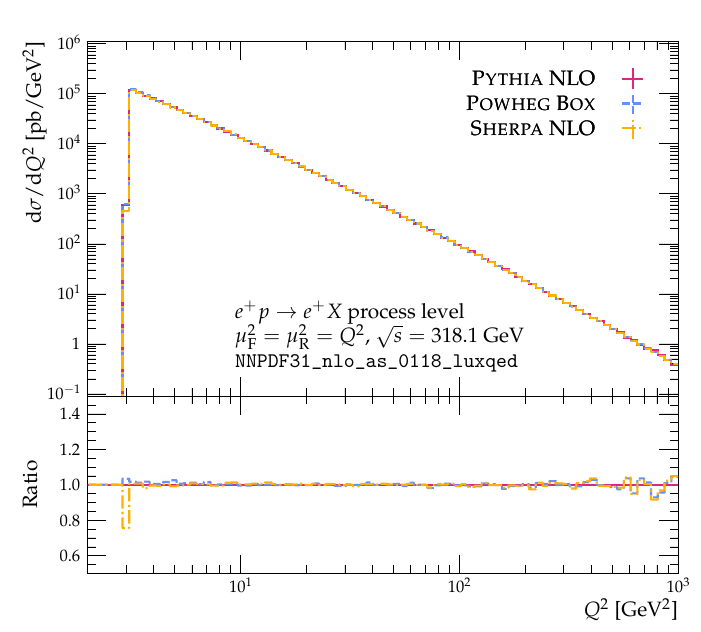}
\includegraphics[width=0.48\textwidth]{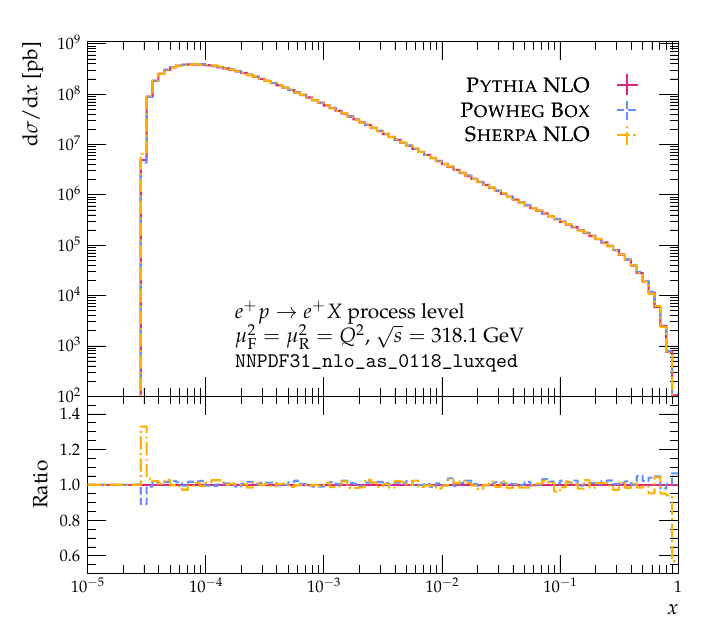}\\
\includegraphics[width=0.48\textwidth]{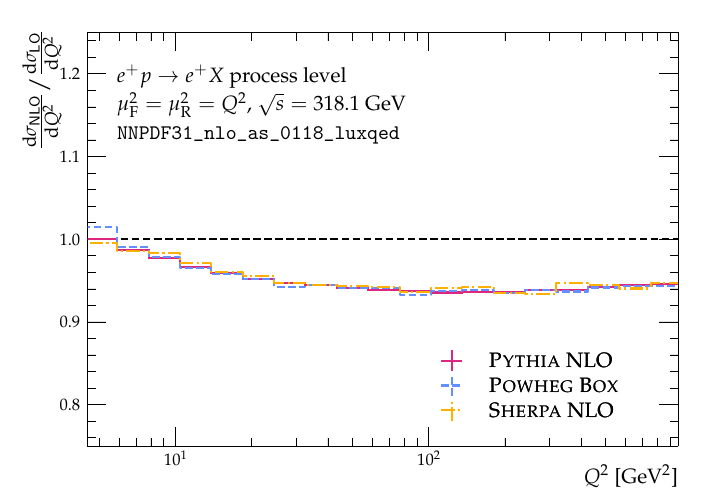}
\includegraphics[width=0.48\textwidth]{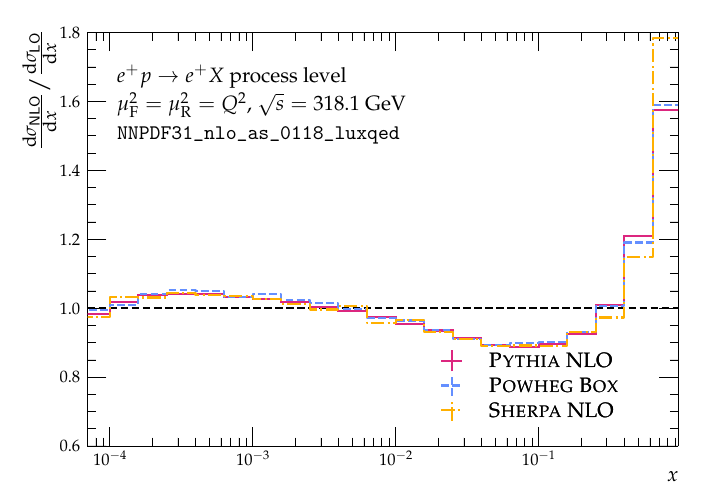}
\caption{Comparison of process-level cross sections (top panels) between different NLO matching implementations and the resulting NLO/LO ratios (bottom panels) as a function of $Q^2$ (left) and $x$ (right).}
\label{fig:Kfactor}
\end{center}
\end{figure}

To validate our implementation of the NLO $K$-factor calculation we compare \textsc{Pythia} process-level cross sections as a function of $Q^2$ and $x$ with the multiplicative NLO matching available in \textsc{Powheg-Box} and the additive matching in \textsc{Sherpa} in figure \ref{fig:Kfactor}. After adjusting all the inputs to the same, we find an excellent agreement with the two other codes for both, the absolute cross section and the NLO/LO ratio for the whole kinematic region. In the comparisons, we have applied NNPDF3.1 NLO proton PDFs \cite{NNPDF:2017mvq} in all the simulations and fixed the factorization and renormalization scales to $Q^2$ and applied a synchronized electroweak parameter scheme.

\begin{figure}[htb]
\begin{center}
\includegraphics[width=0.48\textwidth]{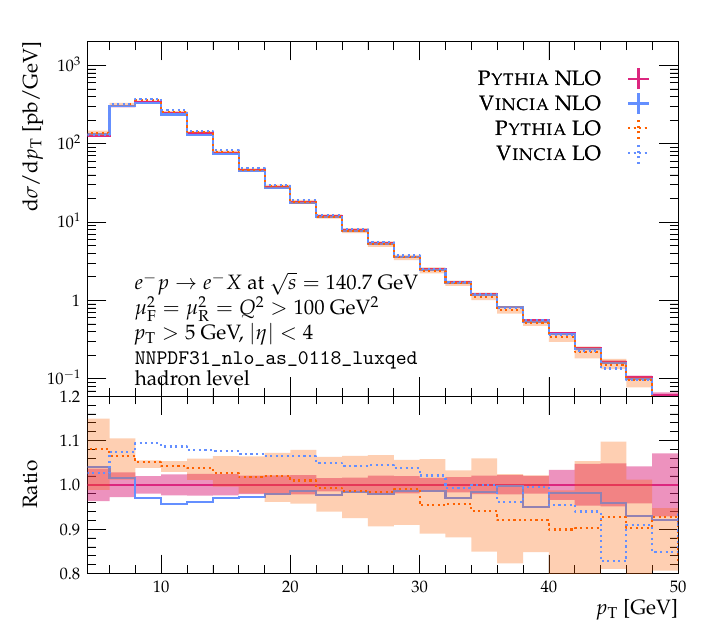}
\includegraphics[width=0.48\textwidth]{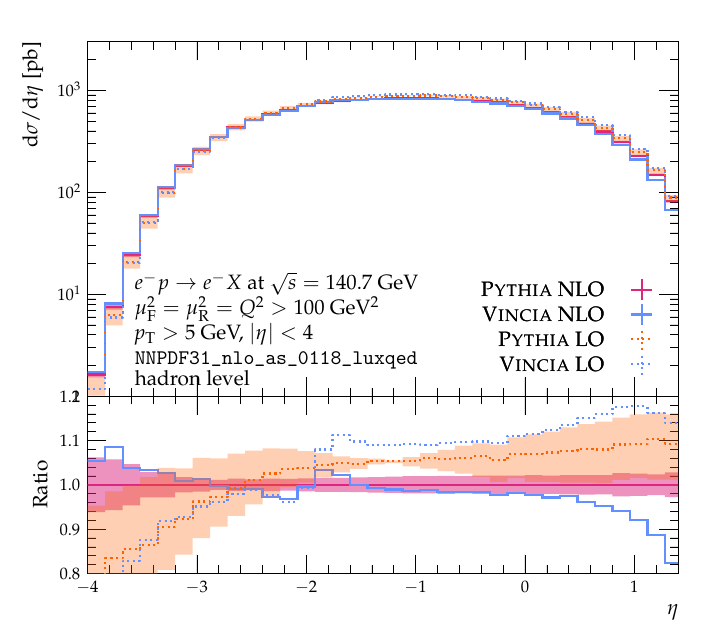}
\caption{Cross section as a function of transverse momentum (left) and pseudorapidity for exclusive 1-jet events at the EIC kinematics with (solid) and without NLO matching (dotted) using the default \textsc{Pythia~8} parton shower with \textsc{dipole-recoil} option (red) and \textsc{Vincia} (blue). Uncertainty bands cover the scale variations which in case of NLO-matched setup includes renormalization and factorization scale variations but only the latter in case LO results.}
\label{fig:eic-jet}
\end{center}
\end{figure}

To study the full impact of the implemented corrections we show the exclusive one-jet particle-level cross sections at the kinematics relevant to the future Electron-Ion collider (EIC) in figure \ref{fig:eic-jet}. Here the simple-shower and \textsc{Vincia} NLO results include both, inclusive NLO $K$-factors and MECs for the first shower emission, whereas LO results are derived without these. There are few observations to be made from the comparisons. First of all, the reduction of scale uncertainties, covering factorization scale and, in case of NLO, also renormalization scale variations, are significantly reduced when NLO corrections are accounted for. We find also that enabling MECs decrease the difference between the two shower implementations, as in the case of NLO-matched results both showers are within $5~\%$ (apart from the extreme kinematics), while without these corrections the difference is found to be somewhat larger.

\section{Comparisons to HERA data}

\begin{figure}[htb]
\begin{center}
\includegraphics[width=0.48\textwidth]{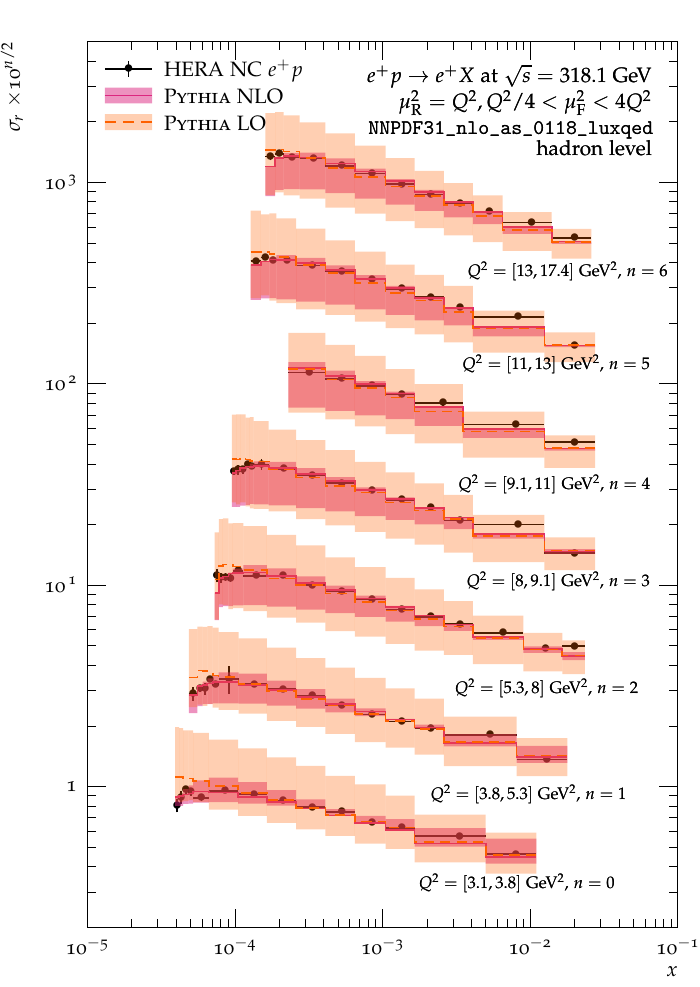}
\includegraphics[width=0.48\textwidth]{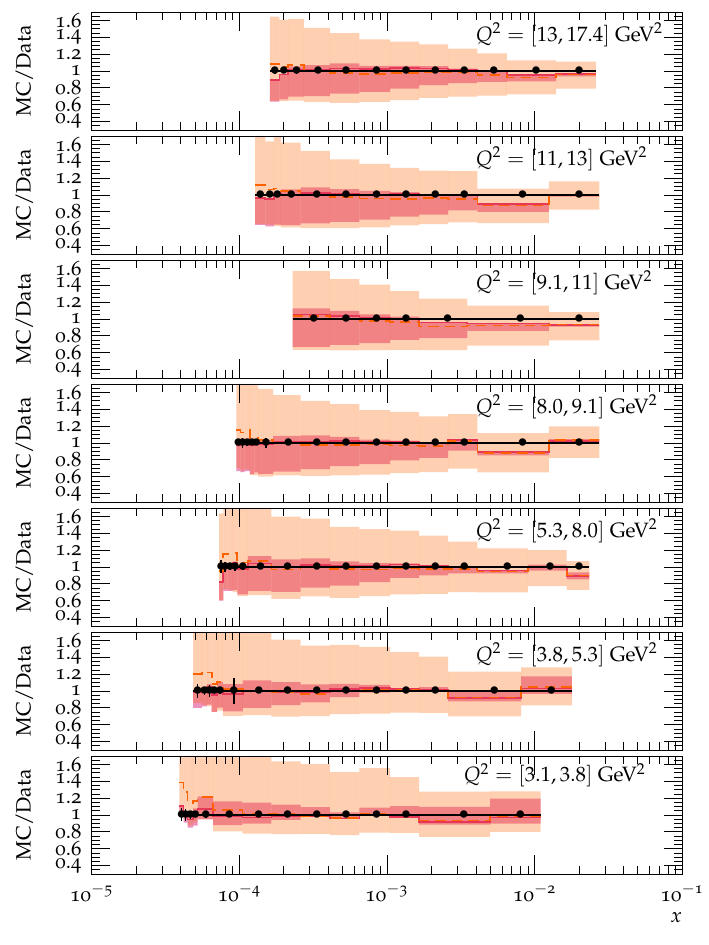}
\caption{Comparison with the reduced cross section data from HERA \cite{H1:2015ubc} with LO (orange-dashed) and NLO-matched (red-solid) results from \textsc{Pythia~8}. Both the absolute cross section (left) and ratio between the data and simulated results (right) are shown for several $Q^2$ bins as a function of $x$. The uncertainty bands show the factorization scale variations.}
\label{fig:sigmaR}
\end{center}
\end{figure}

We present comparisons to the reduced cross section data, $\sigma_r$, compiled from measurements performed by the H1 and ZEUS experiments at HERA \cite{H1:2015ubc} in figure \ref{fig:sigmaR}. To highlight the impact of NLO effects we show particle-level results at LO and NLO including factorization-scale variations.  In addition to the absolute cross section comparison, we also show the ratio between data as a function of $x$ for a subset of $Q^2$ bins at the highest analyzed collision energy. The main observation is again the reduced impact from scale variations and improved description of the data in the small-$x$ region when NLO corrections are included. The full range of $Q^2$ bins are shown in Ref.~\cite{Helenius:2026odp} whereas the selected subset highlights the region where the NLO effects have the largest impact but staying away from low-$Q^2$ region where the validity of DIS description starts to be questionable. The good agreement is of course expected for such an inclusive observable but it serves as an important test nevertheless to validate that the NLO accuracy is retained even when parton showers with the implemented MECs and hadronization effects are included in the simulation. 

\section{Summary and Outlook}

We have presented a new multiplicative matching implementation for neutral-current deep-inelastic scattering processes in the \textsc{Pythia~8} event generator. The implementation is based on an explicit event-by-event calculation of NLO cross sections applying well-known structure functions and matrix-element corrections for the first parton-shower emission to match the real-emission rate in the NLO calculation. We have implemented these corrections for the \textsc{dipole-recoil} option of the default Pythia parton shower and to \textsc{Vincia}, which both keep the scattered-lepton kinematics intact when distributing recoil from QCD emissions. We have validated the implementation against other NLO event generators and experimentally measured cross sections over a broad kinematic range. In addition, we have studied the impact of MECs and NLO structure functions in exclusive 1-jet rates at the kinematics relevant to the EIC. The implementation is the first NLO matching available internally in \textsc{Pythia} for collisions involving hadronic beams. The framework is set up in such a way that future extensions for other processes is relatively straightforward, though the integration over the real-emission phase-space needs to be solved separately for each process. We plan also to include MECs relevant for DIS process into the new next-to-leading-log (NLL) accurate parton shower \textsc{Apollo} \cite{Preuss:2024vyu} which is currently being extended to handle collisions involving hadron beams.

\section*{Acknowledgements}

\noindent This research was funded through the Research Council of Finland, project 361179, and the Center of Excellence in Quark Matter, project 346326, and by the Deutsche Forschungsgemeinschaft (DFG) under grant 396021762 - TRR 257: Particle Physics Phenomenology after the Higgs Discovery. Finnish IT Center for Science, project jyy2580, is acknowledged for computing resources.

\bibliographystyle{JHEP}
\bibliography{Refs.bib}

\end{document}